\documentclass[12pt,dvips]{article}
\usepackage{epsfig}

\begin{document}
%%%%%%%%%%%%%%%%%%%%%%%%%%%%%%%%%%%%%%%%%%%%%%%%%%%%%%%%%%%%%%%%%%%%%%%%%%%
\title{
\vspace{-5.0cm}
\begin{flushright}
{\normalsize WUB 96-30}\\
\end{flushright}
\vspace*{1cm}
\vfill
Multicanonical Study of Continuum Physics in the
D=2 $O(3)$ Nonlinear Sigma Model}

\author
{\bf T.~Neuhaus\\ \\
FB8 Physik, BUGH Wuppertal, Germany}
\date{\today}
\maketitle
\thispagestyle{empty}

\begin{abstract}
Employing a variant of the Multicanonical
Ensemble we study
twisted spin configurations on periodic boxes in the $D=2$ $O(3)$
nonlinear sigma model
for $\beta$-values inbetween $1.55$ to $3.1$. The
free energy difference of twisted spin configurations
is determined
from the constraint effective
potential. The finite size scaling behavior
is in accordance with the asymptotically free nature
of the continuum theory. Upon certain reasonable assumptions we
determine the $\Delta \beta( \beta)$-shift of the
stiffness correlation length $\xi_s$.
The
mass-gap as determined by our analysis is
${m_0}=79.6(1.9)~\Lambda_{latt}$.
This value agrees with the
analytical result of the
thermodynamic Bethe Ansatz
${m_0}=80.1~\Lambda_{latt}$.
\end{abstract}
\newpage

\section{Introduction}

The two dimensional $O(3)$ nonlinear sigma model
is a asymptotically free field theory \cite{a_f}. It deserves special interest in view of
asymptotically free four dimensional nonabelian gauge theories
describing the strong interactions inbetween Quarks. A fundamental
feature of these theories is the existence of a nonvanishing
mass-gap $m_0$. Recently the mass-gap of the sigma model
has been calculated analytically on the basis of the
thermodynamic Bethe Ansatz \cite{HaNi1,HaMaNi1}. Choosing the simplest
regulator, namely a two dimensional
hypercubic lattice with lattice spacing $a$, and the standard nearest neighbor action functional with nearest neighbor coupling $\beta$ we consider
\begin{equation}
 S= - \beta \sum_{<i,j>} \phi_i^{\alpha} \phi_j^{\alpha},
\end{equation}
where the fields $\phi^{\alpha}_i$ are $3$-component unit vectors and the
sum $<i,j>$ runs over the nearest neighbors on the lattice.
The mass-gap correlation length $\xi_0=m_0^{-1}$ has the value
\begin{equation}
\xi_0^{theor}=a \ {e^{{1-\pi/2}}\over{8 \ 2^{5/2}}} \
{e^{2 \pi \beta}\over{2\pi\beta}} \
(1-{.091\over\beta}+O(1/\beta^2)),
\end{equation}
corresponding to the exact mass gap $m_0 = 80.0864~\Lambda_{latt}$ \cite{HaMaNi1},
$\Lambda_{latt}$ denoting the 3-loop lattice cut-off parameter \cite{three_loop}.
The past numerical
studies of the path
integral \cite{HaNi2,Wolff}
consistently overestimated this
number.
These studies considered
infinite volume mass-gap correlation length values
$\xi_0$ up to about hundred lattice spacings $a$
with the help of cluster algorithms. It was also attempted
to enlarge the accessible correlation length region utilizing
real space renormalization group methods \cite{HaNi2}.
It was found, that asymptotic perturbative
scaling only sets in at mass-gap correlation length
values, which can hardly be fitted
into the memory of todays computers.
In addition it may also be
appropriate to mention the still standing
criticism of Patrascioiu, Seiler and others on the continuum limit of
the sigma model \cite{Seiler1}. These authors argue in favor of a
the existence of a phase with vanishing mass-gap in the sigma model, which
is however not supported by the numerical data \cite{o3_sim}.
It is therefore
a challenge to confront the known analytical result on the mass-gap
with the precision numerical simulations of the path integral.
From such a study we might also gain better insight on how to
accurately determine
the mass-gap of lattice QCD.

  Finite size scaling theory \cite{Fisher1} has been very successful
throughout the years in predicting the behavior of the infinite volume
correlation length of spin systems at criticality
from the properties of finite systems, which in this
paper are taken to be square boxes with linear extent $L$ and periodic boundary conditions.
In a
situation where the control parameter $x_0=\xi_0/L$
is large and, even at the critical
point, where the correlation length is infinite, the path integrals
singular part of the free energy on finite volume systems is a function of
the control parameter $x_0$ alone. The
Fisher scaling
analysis leads
to reliable determinations
of the correlation length divergence critical exponent $\nu$ in
statistical mechanics models.

In this paper we attempt a generalization of this idea to the $D=2$ nonlinear
sigma model. Its asymptotic freedom fixed point is located
at infinite value of the bare
nearest neighbor coupling. The quantity under consideration is
a free energy difference of twisted relative to untwisted spin configurations
- the spin stiffness - forming a Bloch wall, whose finite size
scaling behavior is dominated by logarithmic terms
in the control parameter $x_s=\xi_s/L$ in accord with
the perturbative treatment of the path integral in the
one-loop approximation. The quantity $\xi_s$ hereby denotes
the stiffness correlation length.
We will argue, that
a precise extraction of the
$\Delta\beta(\beta)$-shift corresponding to the
stiffness correlation length is feasible
up to $\beta$-values as large as $3.1$, upon
reasonable assumptions on the coefficients of the
finite size scaling law.
The integration of the shift in junction with a start value for the
mass-gap correlation length $\xi_0$
allows the determination of the mass-gap $m_0$.
The paper is organized as follows:
Section 2 is devoted to the theoretical background and introduces
the constraint effective potential, the considered free energy
difference and discusses
the finite size scaling hypothesis. Section
3 contains details of the Multicanonical Ensemble simulation. In section 4 we present
our numerical analysis and results. Section 5 concludes the
of the paper.

\section{Theoretical Considerations}

In statistical field theories, which share the
universality class of continuous ferromagnets, it is well known
on the basis of the $\epsilon$-expansion and the renormalization
group \cite{rudnick_jasnow}, that
the critical behavior of the system
can be studied in the symmetry broken phase of the theory
by considering path integrals of
configurations, which
interpolate inbetween regions of different order parameter orientation.
The helicity modulus $Y$ carries
the information on the nonanalytical behavior in the
symmetry broken phase of the theory. It is common to define the
helicity modulus by the response of the system with respect to a twist angle
$\Theta$, which is applied microscopically
to the fields on the boundary of the lattice
in one of the lattice directions. Denoting
${F=-lnZ}$ the free energy, the free energy difference
\begin{equation}
\Delta F_Y(\theta)= F(\theta)-F(\theta=0)
\label{free_energy_difference}
\end{equation}
then defines the helicity modulus \cite{helicity}
\begin{equation}
Y = \lim_{(\Omega,L) \to \infty }
{{2 \ L} \over { \theta^2 \ \Omega}} \Delta F_Y(\Theta),
\label{helicity_modulos}
\end{equation}
which is finite in the thermodynamic limit and allows for a
geometry independent i.e., $\Theta$-independent,
characterization of the Bloch wall.
Hereby denotes $\Omega$ the cross section orthogonal to the direction
of the twist.
E. g., on a hypercubic lattices with linear extent $L$ one has
$\Omega \propto L^{D-1}$ and one expects $\Delta F_Y(\theta) \propto L^{D-2}$.
The critical behavior of the theory in terms of the helicity modulus
is expressed by Josephsons scaling law \cite{josephsons_scaling_law}:
$Y = A_{Y} t^{\mu_J}$
with $\mu_J=(D-2)\nu$.
It is clear, that
the presented scaling laws refer
to the ferromagnetic
case.
In fact, when the dimension $D=2$ is approached
from above for field theories with continuous symmetry
, like the $O(N)$ nonlinear sigma model with $N \ge 3$ \footnote{We refrain
from a discussion of the $O(2)$ symmetric XY model.},
the ferromagnetic character of the theory is lost
and the helicity
modulus $Y$ according to eq.(\ref{helicity_modulos}) vanishes
for any
value of the bare coupling $\beta$. In dimension $D=2$
the infinite volume system exhibits domains of different order parameter orientations, which
are free to fluctuate.
This is the content of the
Mermin Wagner theorem.

At the heart of the problem is a study of the correspondingly defined free energy difference
$\Delta F_s(\Theta)$ on finite lattices in the $D=2$ $O(3)$ nonlinear
sigma model at large values of the nearest neighbor coupling i.e.,
in a situation in which the mass-gap correlation length $\xi_0$ exceeds the linear
extent $L$ of the symmetric hypercubic lattice. In this situation one expects
$\Delta F_s(\Theta)$ to be controlled by $x_s=\xi_s/L$, where
$\xi_s$ denotes the stiffness correlation length.
Based on the renormalization group, the
$\epsilon=D-2$-expansion and the 1-loop approximation, the
calculation has been performed for the case of a  continuum square box
with fixed twisted boundary
conditions \cite{Chakravarty,Zinn1}.
We quote the result
\begin{equation}
\Delta F_s(\Theta)= {\Theta^2 \over {4\pi}} [ {{ln}(\xi_s/L)}
+{{lnln}(\xi_s/L)} ]
+ R(\Theta)
+ {\cal O(1)}.
\label{zinn_formula}
\end{equation}
A few remarks are in order:

1) According to the expected Fisher
scaling $\Delta F_s(\Theta)$ splits into
a singular part, which is a function of $x_s={{\xi_s}/{L}}$ alone
and a regular contribution $R(\Theta)$.
The stiffness correlation
length $\xi_s$ follows the perturbative renormalization group, and
thus in the large $\beta$-limit has the form
\begin{equation}
\xi_s(\beta) \propto  {1\over\beta}e^{2 \pi \beta}.
\label{asymptotic_scaling}
\end{equation}
However, $\xi_s$ is not identical to the
mass-gap correlation length $\xi_0$ as is the control parameter
$x_s$ as compared to $x_0$.
The stiffness correlation length $\xi_s$
describes the crossover of the system from its small volume behavior
at large values of $x_s$
into a state of various domains
of order parameter orientation on large volumes. It is expected
to be larger than the mass-gap correlation length $\xi_0$.
Recently Billoire performed a high statistics numerical calculation
of the stiffness correlation length \cite{Alain} at $\beta$-values
far in the perturbative region
$\beta =5,10$ and $20$ respectively. Expressing the measurement
in units of the exact mass-gap correlation length we quote
$\xi_s = 9.39(1), 9.46(1)$ and $9.47(1) \xi_0^{theor}$.
In this paper we assume
$\xi_s$ takes its perturbative value
that at a $\beta$-value of $3$.
In addition we expect
a region of $\beta$-values, where $\xi_s$
as well as $\xi_0$ are governed by the same
beta-shift $\Delta\beta(\beta)$. Let us denote with $\Delta \beta(\beta)$
the change of the coupling $\beta$ to $\hat{\beta}=\beta-\Delta\beta(\beta)$
corresponding to the decrease of the stiffness correlation length by a factor
of $2$ : $\xi_s(\hat{\beta})={1\over2}\xi_s (\beta)$.

2) The free energy difference
$\Delta F_s(\Theta)$ at $\Theta=0$ defines the $\Theta$ independent spin stiffness
$\rho={2 \over \Theta^2} \Delta F_s(\Theta)$. The large $\beta$
perturbative limit of $\rho_s=-{\partial\over{\partial lnL}} \rho$
defines the spin stiffness constant whose value
is ${1\over{2\pi}}$. Precision numerical simulations of the spin
stiffness \cite{Alain,Caffarel} confirm
this theoretical expectation.
In the context of the present paper we assume
that the spin stiffness coefficient proportional to $ln L$
takes its perturbative value in the whole considered $\beta$ region.

3) Upon insertion of eq.(\ref{asymptotic_scaling}) into
eq.(\ref{zinn_formula}) we observe, that the leading term
in $\Delta F_s$ is
${\beta  \Theta^2 / {2}}$. This is the classical
action difference of a twisted configuration at twist angle $\Theta$
relative to a configuration with twist angle
$\Theta=0$. The dominant effects of quantum fluctuations
at fixed $\beta$
are linear in $lnL$
and are given by a term
$-{\Theta^2}{lnL}/4\pi$. They drive the spin stiffness
to smaller values on the larger system sizes, as expected.

4) The regular contribution $R(\Theta)$ in eq.(\ref{zinn_formula})
can be calculated in the small $\Theta$ expansion. For the
boundary conditions considered in \cite{Zinn1} one finds \cite{Alain}
$R(\Theta)={-2.501{{\Theta^4}\over{8\pi^4}}+{\cal O}(\Theta^6)}$ i.e., a negative
contribution at finite $\Theta$ values. We note that these
contributions are nonuniversal in character and depend on the details
of the implementation of the twist.

Numerical simulations of free energies are computationally
difficult. In the standard approach one integrates the
expectation value of the action difference of systems with nonzero twist $\Theta$
and twist $\Theta=0$ along the nearest neighbor coupling parameter direction $\beta$.
One can avoid
the integration of the action
by differentiating the path integral
with respect to $\Theta$ at a value of $\Theta=0$ \cite{Alain,KKMon}.
In this paper we consider the constraint effective potential
(CEP) \cite{CEP_1} of the mean field of the theory.

The consideration of the CEP is motivated by the
analogous problem arising in the interfacial case in $Z(2)$ symmetric theories and
recently numerical simulations of it have been conducted with the
help of Multicanonical Ensemble simulations \cite{Berg_Neuhaus}.
One considers the mean field
$ M= {1\over L^D} \sum_i S_i $ of the Ising fields.
Its probability distribution
$P(M)$
contains at the same time the information concerning
bulk as well as
interfacial properties of the system.
Relevant for interfacial effects
are states at $M=0$
on finite boxes with periodic boundary conditions.
In this region of the phase space
configurations contain two domains of opposite order parameter
orientation and two interfaces are formed.
Widoms scaling
law can be
analyzed \cite{OurIsing}.
Lee and
Kosterlitz have performed a
finite size scaling analysis of the CEP at criticality
\cite{LeeKosterlitz} leading to a determination of $\nu$.

In case of the theory with continuous $O(3)$ symmetry
we introduce the $3$-component field
\begin{equation}
\Phi^{\alpha}= {1 \over L^2} \sum_x  \phi_x^{\alpha}
\end{equation}
and its absolute value, denoted the mean field $\bar{\Phi}$ in the following
\begin{equation}
{\bar{\Phi}} =\sqrt{ \sum_\alpha \Phi^{\alpha}  \Phi^{\alpha} },
\end{equation}
which
will
have the probability distribution
\begin{equation}
P({\bar{\Phi}}) \propto {{\bar{\Phi}}^2} e^{-U(\bar{\Phi})}
\label{P_can}
\end{equation}
in the canonical ensemble of the theory.
The function $U(\bar{\Phi})$ is the CEP
of the theory. It can be
obtained by rewriting the partition function
\begin{equation}
Z=\int D\phi e^{\beta \sum_{x,\nu} \phi_x^{\alpha}\phi_{x+\nu}^{\alpha}}
\label{the_path_integral}
\end{equation}
into
\begin{equation}
Z=\int_0^\infty d{\bar{\Phi}} {\bar\Phi}^2 e^{-U(\bar\Phi)},
\end{equation}
which can be achieved by introducing suitable $\delta$-functions
and integration upon the remaining degrees of freedom.
We note that $U(\bar\Phi)$ is defined up to constant, which can
easily be absorbed into a multiplicative normalization of $Z$.
In addition a factor ${\bar\Phi}^{N-1}$ at $N=3$ appears in the path integral definition
of $U$. This factor is proportional to the surface of a
$N$-sphere of radius $\bar\Phi$ and accounts for the degeneracy
of states with respect to $O(N)$ rotations.
Without taking this phase factor
properly into account
the CEP
is a singular function at $\bar\Phi=0$.

The CEP attracted in the past attention in the context
of the Higgs-models represented by $O(N)$ symmetric
theories in $D=3$ and $D=4$.
There it was studied
with analytic methods \cite{Leutwyler_Goeckeler} as well
with numerical simulations \cite{Neuhaus_et_al}.
In these models the CEP has at a finite value $\bar\Phi_{{min}}$
a minimum, which corresponds to the Higgs field expectation value.
We may note, that both analytic, as well as the numerical
considerations in these models were concerned with the shape
of the CEP in the vicinity
of its minimum.

In the context of the present paper we are however concerned
with the CEP at values of the mean field $\bar\Phi$ equal to
$\bar\Phi=0$. We argue that states at
$\bar\Phi=0$ correspond on hypercubic boxes with periodic boundary
conditions to Bloch walls carrying formally a
twist angle of $\Theta=2\pi$.
This can be easily seen on the classical
level by noting, that the field configuration
\begin{equation}
\phi^{1} = cos({{2\pi n_1}\over{L}}) ~~~
\phi^{2} = sin({{2\pi n_1}\over{L}}) ~~~
\phi^{3} = 0
\label{the_classical_state}
\end{equation}
minimizes the action functional for the given periodic boundary
conditions at $\bar\Phi=0$. This configuration exhibits
on the scale of one lattice spacing $a$ spin twists of value
${2\pi} \over L$ in the 1-direction of the lattice, which upon
integration in the 1-direction adds up to a total twist angle of $\Theta=2\pi$.
We expect this configuration
and the added quantum fluctuations
to dominate
the $\bar\Phi=0$ state.
Thus
in the framework of the CEP it is strongly suggested,
that the singular part of $\Delta F_s(\Theta=2\pi)$
exhibits the
same finite size scaling as the constraint
effective potential difference, or potential barrier
\begin{equation}
\Delta U=U(\bar\Phi=0)-U(\bar\Phi_{min})=\Delta F_s(\Theta=2\pi)+\hat{R},
\label{hypothesis}
\end{equation}
up to regular terms $\hat R$, provided the CEP exhibits
a maximum and minimum corresponding to the twisted and untwisted states.
It is this finite size scaling hypothesis, which we will examine
in the subsequent analysis of this paper. We will find, that the
numerical evaluation
of the CEP confirms the hypothesis.

\section{The Multicanonical Ensemble Simulation}

By the use of standard Monte Carlo algorithms
it would be very difficult to sample
values of $\bar\Phi$ close to zero, if
the value of the nearest neighbor coupling $\beta$ is larger
than about $1.6$ on lattices of reasonable linear size $L$.
To overcome this difficulty, we modify the importance
sampling by introducing a Multicanonical-weight
factor into the Monte Carlo sampling process \cite{Berg_Neuhaus}.
We remark, that the idea of Multicanonical Ensemble
simulations consists in modifying the Boltzmann-weight for the
purpose of the improvement of the actual Monte Carlo sampling process.
The modification is however done in
such a way, that the effect of the Multicanonical-weight
will be removed in a well
controlled way. Thus in the end the CEP
can be determined in the canonical ensemble of the theory.

In case of the CEP the Multicanonical-weight factor is chosen to be
a function of the mean field $\bar\Phi$ evaluated on each
single configuration and will be denoted
$W_{mc}(\bar\Phi)$. It is in principal defined on the real interval
$[0,L^2]$ and it is sensible to split the Multicanonical-weight
into a part, which takes care of the degeneracy with respect
to $O(3)$ rotations, and a yet unknown contribution
$\hat W_{mc}(\bar\Phi)$. Thus for the purpose of the Monte Carlo
simulation we consider the Multicanonical-weight
\begin{equation}
W_{mc}({\bar\Phi})=-2~ln{\bar\Phi}+{\hat W}_{mc}({\bar\Phi})
\label{the_weights_1}
\end{equation}
and
\begin{equation}
e^{-S+W_{mc}(\bar\Phi)}
\label{the_weights_2}
\end{equation}
to be the Boltzmann-factor, which generates the Markov process.
In order to obtain a representation of the
Multicanonical-weight $\hat{W}_{mc}$, which in a
computer is easily calculable, we choose $\hat W_{mc}$ to be a polygon, which
on a finite set of $i=1,m$ intervals $I_i:\bar\Phi_{i}\le\bar\Phi<\bar\Phi_{i+1}$
for each interval is characterized by two parameters $g_i$ and $h_i$:
\begin{equation}
\hat{W}_{mc} (\bar\Phi)= g_i+h_i \bar\Phi.
\end{equation}
One may view the parameters $h_i$ as
magnetic sources, which are applied on piecewise intervals
of the operator $\bar\Phi$ to the theory. In one
of our earlier publications we therefore named the resulting
ensemble the Multimagnetical Ensemble \cite{OurIsing}.
We found it sufficient in the actual numerical
simulations, to first determine a value $\bar\Phi_{max}$ in a
standard simulation. In this situation
$\bar\Phi_{max}$ is calculated in such a way, that all the relevant
structure of the CEP, namely the location of its minimum, and the
states at $\bar\Phi=0$ are included
in the considered $\bar\Phi$-interval. It means that we will only evaluate
the shape of the CEP for values $0\le\bar\Phi\le\bar\Phi_{max}$.
We then choose for reasons of simplicity an aequidistant
partition of the considered $\bar\Phi$-interval into
$m=25$ or $m=40$ different intervals $I_i$, which in the actual
simulations turned out to be a sufficiently fine polygon
approximation to the Multicanonical-weight $\hat{W}_{mc}$.

Multicanonical Ensemble simulations are defined by the requirement,
that the considered operator exhibits an almost constant, ideally
a constant,
probability distribution in the
simulation. If in the canonical ensemble the density of states
function with mean field $\bar\Phi$ is denoted $n(\bar\Phi)$,
then the ideal Multicanonical-weight factor is
\begin{equation}
W_{mc}(\Phi_0)={-ln~n(\bar\Phi)}.
\end{equation}
Though trivial, we note, that a priori the Multicanonical-weight
is not known. Less trivial
we remark, that it is possible to obtain an estimate of the weight
factor from the Monte Carlo simulation. In the vicinity
of a given value of the operator $\bar\Phi$ one may sample the phase space
in an attempt to estimate the density of states. In practice
we perform for each of the considered intervals $I_i$
a separate simulation with $\bar\Phi$-values constrained to the given interval.
In the simulation each parameter $h_i$ is then determined
under the requirement that
the probability distribution of the mean field $\bar\Phi$
approximates the Multicanonical distribution at its best.
The full set of all parameters $h_i$ serves
the purpose of the construction of the complete Multicanonical-weight
according to eq.(\ref{the_weights_1}) and eq.(\ref{the_weights_2}). Clearly, for given $h_i$ the
parameters $g_i$ can be chosen in such a
way, that the resulting polygon is a steady function of $\bar\Phi$.
We note, that one parameter value, e.g. $g_1$ is left undetermined by this
procedure. This again corresponds to the overall normalization freedom
of the path integral.
In our actual simulations
we have found, that this simple and robust procedure results into acceptable
Multicanonical-weights and distribution functions.

Once the Multicanonical-weight factor has been determined, we
implement one sweep of the Monte Carlo sampling
by a combination of a Swendsen-Wang
reflection cluster update \cite{SweWa,Wolff_c}, followed by a subsequent accept-reject
decision, which then depends
on $W_{mc}$ alone. It means, that
we define the Cluster degrees of freedom from the cinetic
term of the action, and, that we replace the usual equal probability
rule of the Cluster update, by a $4$-hit Metropolis accept-reject
decision in all the Cluster degrees of freedom, depending on the magnetic properties of the system. At this
value for the number of hits we obtained average acceptance rates for moves
of the system at about one half for almost all the considered $\beta$ and $L$ values.
In order to monitor the performance of this update procedure, we divided
the considered $\bar\Phi$-interval into an arbitrarily chosen $8$, but equally spaced
bins. We measure the average flip-autocorrelation time $\tau_{f}$ for a transition
of the system from the first to the eights bin, or vice versa.
While due to the partitioning
some arbitrariness in the definition of $\tau_f$ is
induced, we nevertheless expect, that it will exhibit the
basic aspects of the Monte Carlo time dynamics. Fig. 1) contains
data for the quantity $\tau_{f}$
obtained on a $L=36$ lattice for $\beta$-values in between
$1.6$ and $3.0$ (circles), and for a fixed $\beta$-value $\beta=2.4$
on lattice sizes ranging in between $L=20$ to $L=68$ (triangles).
One notices a typical time scale of several thousand sweeps, which
in general is characteristic for our simulation. In detail we observe, that
at a given $\beta$-value the flip-autocorrelation time
stays almost constant, or even decreases with increasing
lattice sizes. We attribute this favorable property to the
nonlocal nature
the algorithm.
We also observe
on the given lattice a rapid increase of
the flip-autocorrelation time with
increasing $\beta$. Such an increase is expected, as
the simulational complexity of the theory will rise in the proximity
of the asymptotic freedom fixed point.

 In the Multicanonical simulation
we measure the Multicanonical probability
distribution $P_{mc}(\bar\Phi)$ of the mean field $\bar\Phi$.
It is related to the probability distribution
in the canonical ensemble $P(\bar\Phi)$
by a simple reweighing step
\begin{equation}
P(\bar\Phi) \propto P_{mc}(\Phi_0)e^{-{W}_{mc}(\Phi)}
\end{equation}
and thus with the help of eq.({\ref{P_can}}) the CEP can be determined
in the Multicanonical simulations. Throughout our simulations we have
fixed
the minimum value of the CEP to be zero.
In addition we perform a bias corrected jackknife error analysis
for all the measured quantities. The typical statistics accumulated
for each data point in our simulation was inbetween $1$ and $5$
megasweeps, depending on the considered $\beta$-value.

\section{Numerical Analysis and Results}

The mass-gap correlation length $\xi_0$ of the $D=2$ $O(3)$ nonlinear
sigma model has been calculated in previous studies at not too
large values of $\beta$ and we
quote here the results of Wolff \cite{Wolff}: At the $\beta$-values
$\beta=1.6,1.7,1.8$ and $1.9$
$\xi_0/a$ takes the values $19.07(06),34.57(07),64.78(15)$
and $121.2(6)$ respectively. In this paper we study
small systems of sizes ranging in between $L=8$
up to $L=82$. For $\beta$-values larger than
$\beta=1.6$ the mass-gap correlation length $\xi_0$ is comparable
with the considered lattice sizes. Accordingly it is expected
that the stiffness correlation length $\xi_s$ exceeds
the linear size of the systems and the onset Fisher
scaling may be expected.
In a first set of simulations
we have studied the $L$-dependence of the CEP
at twelve values of the nearest neighbor coupling, namely
at the $\beta$-values $\beta=1.6,1.7,1.8,1.9,2.0,2.1,2.2,2.3,2.4$
and at the $\beta$-values $\beta=2.6,2.8$ and $\beta=3.0$.
In a second set of simulations the $\beta$-dependence of the CEP
potential was determined on a $L=36$ lattice at $\beta$-values
ranging inbetween $\beta=1.55$ up to $\beta=3.1$. In total we have
accumulated $468$ pairs of couplings $\beta$ and
lattice sizes $L$ within our simulation.

It is important to check our basic assumption, that states at
mean field $\bar\Phi=0$ carry a twist angle of $\Theta=2\pi$.
For this purpose we use a special graphical representation
of the field configuration, which is exhibited in Fig. 2a)
and Fig. 2b).
On the given configuration
we
apply in a first step a global
rotation which equals the arbitrarily chosen
$3$ component of the field $\Phi^\alpha$ to zero.
We then map the $1$ and $2$ components of the fields along paths
in the lattice onto the complex plane. In this mapping we connect
fields, which are neighbors within the path, by a line.
Fig. 2a) contains all such mappings, which are obtained
by considering all $L$ paths along the main axis of the lattice in the
1-direction. Fig. 2b) contains the analogous graphs
for paths along the main axis in the 2-direction.
The considered lattice size is $L=36$ and the nearest neighbor
coupling is $\beta=8$, far in the perturbative region. The
value of $\bar\Phi$ for the considered
configuration was $\bar\Phi=0.003904$, which is very close to zero.
One clearly observes the winding
angle of $2\pi$ for the latter case i.e., for all paths on the
main axis of the lattice in the 2-direction, which
is clear evidence in favor of the above assumption.
It is noteworthy to remark that fluctuations around the classical
state eq.(\ref{the_classical_state}) are sizable even for such a large $\beta$-value. They
increase as $\beta$ is lowered and decrease as $\beta$ is enlarged.
In addition we also mention, that
states at $\bar\Phi=0$ apparently break the cubic group
lattice symmetry i.e., in the here considered configuration
the twist appears in the 2-direction of the lattice.
We have checked, that statistical independent configurations
at $\bar\Phi=0$ assume the possible different twist directions
with equal probability. From a theoretical point of view
it must be expected that this
additional degree of freedom, as compared
to a theory with a fixed direction of the twist, contributes to
the unknown regular contributions
of the finite size scaling relation eq.(\ref{hypothesis}).

In Fig. 3a) and Fig. 3b) we display the CEP, as obtained by our
simulations on lattice sizes $L=24,36$
and $L=70$ and at a values of $\beta=1.6$ and $\beta=2.4$.

At $\beta=1.6$
we observe
a crossover from a situation, where for the smaller $L$-values
the CEP for states at $\bar\Phi=0$
exhibits a barrier $\Delta U > 0$, to a situation at the largest $L$-value,
where the barrier vanishes. Thus the decrease of the
spin stiffness with increasing $L$ is observed.
This is attributed to the quantum fluctuations around the classical
state with twist angle $\Theta=2\pi$. We may remark, that
in simulations of ferromagnets, namely the $D=3$ $O(3)$ sigma model in its symmetry
broken phase, we have witnessed an increase of the potential barrier
with increasing $L$.

At $\beta=2.4$ the mass-gap correlation length and $\xi_s$ exceed
the linear lattice size by a large factor. Correspondingly
a crossover cannot be observed and
$\Delta U \approx 20$ is finite
on the considered range of lattice sizes. Again
its decrease with increasing $L$ is witnessed.
The large value of $\Delta U$ corresponds
to an exponentially large suppression of states with
mean field $\bar\Phi=0$, even if the phase space factor
of eq.(\ref{P_can}) is divided out from the probability distributions.
In fact, a simulation of the $D=2$ $O(3)$ nonlinear sigma model at $\beta=2.4$
on small volumes resembles, as far as order parameter orientation
is concerned, on the first sight the behavior of a ferromagnetic system.
It is only, that on lattices with linear extent comparable
and larger to $\xi_s$ , the system will approach
its true vacuum state.

From the shape of the CEP we have carefully determined
the quantity $\Delta U=U(\bar\Phi=0)-U(\bar\Phi_{min})$.
For this purpose we employ fits to the shape of
the CEP in the vicinity of the minimum at $\bar\Phi_{min}$
and in the vicinity of $\bar\Phi=0$.
In the vicinity of $\bar\Phi=0$ and for the determination
of $U(\bar\Phi=0)$ we describe the CEP by a parabola. Such
a functional form can be expected on the basis
of a classical argument
\begin{equation}
U= U(\bar\Phi=0)+\alpha_1{\bar\Phi}^2
\end{equation}
and is consistent with the data.
We use it in a fit range of $\bar\Phi$-values close to $\bar\Phi=0$, for
which corresponding $\chi^2_{d.o.f}$-values of the fits are smaller than unity.
Inspecting the shape of the CEP in vicinity of its minimum
we note an apparent asymmetric behavior and therefore
we use the form
\begin{equation}
U= U(\bar\Phi_{min})
 +\beta_1 (\bar\Phi-\bar\Phi_{min})^2
 +\beta_2 (\bar\Phi-\bar\Phi_{min})^3
\end{equation}
for its analytic description and the determination
of $U(\bar\Phi_{min})$. Analogous remarks as in the previous case
on the fit-intervals and the $\chi^2_{d.o.f.}$-values apply.

We expect, that
discretization effects of the lattice
theory, as compared to the continuum, lead to correction terms
to the spin stiffness. The
leading contributions are of the order $1/a^2$.
On the lattice and in the classical approximation we calculate
the action difference $\Delta S_{0,latt}$ of a field configuration with
twist $\Theta=2 \pi$ relative to untwisted fields. It has
the expansion
\begin{equation}
\Delta S_{0,latt}= \Delta S_{0,c}[\rho(L)+{\cal O}({1\over a^4})],
\end{equation}
with $\rho(L)$ given by
\begin{equation}
\rho(L)=1-C_g{\pi^2\over{3(La)^2}}.
\end{equation}
Classically the constant $C_g$ is equal to unity, and
$\Delta S_{0,c}=2\pi^2\beta$ denotes the action difference of the Bloch wall
in the continuum.
We will assume here, that the possible form
of the $1 / a^2$-corrections
is already determined on the classical level. Correspondingly
we assume in the fully fluctuating theory, that
additional contributions can be accounted for by a nonunity value
of the parameter $C_g$. The potential barrier in the continuum
$\Delta U_c$ is then related to $\Delta U$ evaluated on the
lattice via the relation
\begin{equation}
\Delta U_c= \rho(L)^{-1} \Delta U,
\label{ctogrid}
\end{equation}
which then allows the comparison of the lattice data with the continuum scaling
form.

We have analyzed the finite size scaling of almost all of our
$\Delta U$ data by means of one $\chi^2$-fit with the form
\begin{equation}
\Delta U=[1-C_g{\pi^2\over{3(La)^2}}][\pi {{ln}(\xi_s/L)}
+ A {{lnln}(\xi_s/L)}
+ \tilde{R}].
\label{final_fit_formula}
\end{equation}
Aiming at a precise
determination of the mass-gap $\xi_0$ we implement our knowledge
about the perturbative limits of the spin stiffness $\rho$
and the spin stiffness correlation length $\xi_s$.
We constrain
certain parameters and adopt
the following scheme:

1) In accord with theoretical expectations
the
prefactor of the single logarithmic scaling term
$\propto \ln(\xi_s/L)$ of the continuum free energy is fixed to its exact value $\pi$. This
is expected on the basis of our previous discussion on the spin twist
and the spin stiffness.
We leave the prefactor $A$ of the double logarithmic
scaling term to be a free parameter. It may be tolerated, that some
of the omitted ${\cal O}(1)$ higher order terms of the loop expansion in eq.(\ref{final_fit_formula})
may be represented effectively by a value of $A$, which differs from $\pi$.

2) The quantity
$\partial_{\beta}ln\xi_s(\beta)$ is expanded
in inverse powers of the nearest coupling parameter $\beta$.
Thus
\begin{equation}
\partial_{\beta}ln\xi_s(\beta)=2\pi-{1\over\beta}+{0.091\over\beta^2}
+ \sum_{k=3}^{6}\gamma_k {1\over\beta^k}
\label{fit_ln_xis}
\end{equation}
represents an analytic form, which incorporates the known 3-loop
behavior of the mass-gap correlation length. It is valid, if
in the asymptotic scaling region $\xi_0$ and $\xi_s$ follow the same
$\beta$ function.
Without loss of generality we have included 4 free parameters
$\gamma_k$ for $k=3,...,6$. These additional degrees of freedom
allow for deviations from perturbative
scaling in the small $\beta$-value region, where the
dip in the $\Delta \beta (\beta)$ function is expected.

3) Given
a start value
for the stiffness correlation length at a value
of the nearest neighbor coupling $\beta^*$
in the perturbative region of the theory, $\xi_s$
can be integrated
to smaller values of $\beta$ via
\begin{equation}
ln\xi_s(\beta)=\int_{\beta^{*}}^{\beta}
\partial_{\beta} ln \xi_s(\beta)  + ln \xi_s(\beta^{*}),
\end{equation}
if within the fit the quantity
$\partial_{\beta}ln\xi_s(\beta)$ is represented by its
expansion eq.(\ref{fit_ln_xis}).
We exploit the recent measurements of the spin stiffness correlation length in the
asymptotic scaling region \cite{Alain} and fix the
integration constant $ln\xi_s(\beta^*)$ by choosing
$\beta^*=3$ and $\xi_s(\beta^*)=9.47\xi_0^{theor}(\beta^*)$.
Note then that at $\beta^*=3$, $\xi_s$ has a value
of $934051a$. Any significant deviation of $\xi_s$
from its measured value in the asymptotic
scaling region would appear unreasonable to us at such
large values of the correlation length.

4) We also express the functional form for the quantity $C_g(\beta)$
through an expansion in inverse powers in $\beta$
\begin{equation}
C_g(\beta)=1
+ \sum_{k=1}^{4}\delta_k {1\over\beta^k},
\label{fit_c_g}
\end{equation}
upon introduction of 4 free parameters $\delta_k$ with $k=1,...,4$
parameterizing the departure of $1/a^2$ corrections from the classical
result. At infinite value of $\beta$ the classical result is recovered.

5) The onset of finite size scaling is expected for large
values of the control parameter $x_s>>1$. Into the actual fit
we include data from the $\beta$-interval $1.65 \le \beta \le 3.1$,
leaving us with $428$ independent data points.
For these data and for the considered lattice sizes our final
fit only includes data, whose
control parameters obey the inequality $x_s>4.7$.
Such large values of the control parameter $x_s$ hopefully
put us in a region of the theory, where we can reliably trust our finite
finite size scaling ansatz eq.(\ref{final_fit_formula}).

The resulting $10$ parameter $\chi^2$-fit for the parameters
$A,\tilde{R},\gamma_k$ and $\delta_k$ with $k=1,...,4$ has been executed
with the double precision version of the CERN libraries MINUIT
program. The error analysis for the parameter values and for all
other derived quantities of this paper has been performed
by a repeated execution of the fit on several data sets. In each data
set any datum in the fit is distributed gaussian around its mean
with a variance corresponding to its error.
The fit has an excellent $\chi^2_{d.o.f}$-value
of $\chi^2_{d.o.f}=0.75$ and is displayed in Fig. 4) and Fig. 5)
in comparison with the measured
$\Delta U$ values. The fit as denoted by the lines in the figures captures the
$\beta$-dependence of the data in Fig. 4) as well
as the $L$-dependence displayed in Fig. 5).
As expected by the theoretical
reasoning the fit supports a negative regular contribution
to the scaling law eq.(\ref{final_fit_formula})
$\tilde{R}=-3.45(20)$. It can be compared with the mentioned result
of the small $\Theta$-expansion of ${\cal O}(\Theta^4)$ on boxes with fixed boundary
conditions at a formal value of the twist angle $2\pi$, which is
$-5.002$. However there is no theoretical argument, that both numbers
should be identical, rather they should differ by contributions
induced by the different boundary conditions.
We also find a positive coefficient $A=1.79(7)$ attached to the double
logarithmic scaling terms. The former value differs from the expected
expected value $\pi$, which we attribute to the
subleading nature of the double logarithmic terms. In Fig. 6) the fit result
for the constant $C_g$ is given.
Finite grid effects
are largest for $\beta$-values around $\beta=1.9$ and
exceed their classical value in the constant $C_g$ by a factor of about $3$ there.
With increasing $\beta$ a very slow approach to the classical result
is witnessed. In Fig. 7) the logarithm of the stiffness correlation length
as well as its $\beta$-derivative are displayed. Note again that we
have fixed $\xi_s$ at $\beta^{*}=3$ to the measured perturbative
result, the fat circle in the figure.
The calculated $\beta$-derivative of $ln\xi_s$ as displayed
by the solid curve in the inlay of Fig. 7)
exhibits a very clear "peak like" deviation from 3-loop asymptotic perturbative
scaling, the dashed curve in the inlay of Fig. 7).
The peak corresponds
to the dip in the $\Delta \beta(\beta)$-shift as
noticed in the earlier studies of the sigma model. It is located
at a $\beta$-value of about $\beta \approx 1.75$. Notably
our data demonstrate, that deviations from 3-loop asymptotic
scaling are present up to $\beta$-values of
about $\beta=2.5$, while for all larger values of $\beta$
asymptotic perturbative scaling is realized within the numerical
precision.
This finding is a genuine outcome of the present analysis
and also applies to the
$\Delta \beta(\beta)$-shift corresponding
to the decrease of the stiffness correlation length by a factor $2$.
The $\Delta \beta(\beta)$-shift is displayed in Fig. 8) for $\beta$-values
larger than $\beta=1.8$, the solid curve in the figure.
It is compared with the 3-loop asymptotic
perturbative scaling result, the dashed curve in the figure and
with results from Hasenfratz et. al \cite{HaNi2}.
Table 1) collects few entries for the shift at selected values of $\beta$.
We thus observe a somewhat slower approach to
asymptotic scaling as it was previously anticipated in the
the numerical simulations. A similar observation has already been
reported earlier \cite{my_muca_o3}.

The calculation of the mass-gap of the theory in units
of the 3-loop lattice cut-off
can be attempted under the nontrivial assumption, that
mass-gap as well as stiffness correlation length
share a common $\beta$-value region, in which their
flow towards the asymptotically free fixed point with varying
couplings is governed by a universal
$\Delta \beta(\beta)$-shift.
In this scaling region the ratio
$\xi_0/\xi_s$ stays constant. It is common believe, that
this region stretches beyond the
region where asymptotic scaling is observed.
Providing a start-value of the mass-gap correlation length
at a small value of $\beta$ outside
the perturbative scaling region, but within the so called nonasymptotic scaling region, it
may be feasible
to integrate the mass-gap of the theory
with the help of
$\partial_\beta \xi_s(\beta)$
up to a $\beta$-value $\beta^*$ in the asymptotic scaling region.
The existence of a such a universal behavior beyond the asymptotic regime
is however by no means
guaranteed.
Referring to
the results of the preceding paragraph we are confident now that
the choice $\beta^*=3$ moves us into the asymptotic scaling region.
For the start value of the mass-gap correlation
length we choose one of Wolffs numbers
\begin{equation}
\xi_0(\beta=1.8)=64.78(15)a.
\end{equation}
This result has been obtained on a quite sizable
$512 \times 512$-lattice with the help of the cluster
algorithm and represents one of the most reliable mass-gap
correlation length measurements at large values of the
correlation length. It is consistent with another work \cite{HaNi2}.
In Fig. 9) we display the integration
of the mass-gap in units of $\Lambda_{latt}$ from its start value at $\beta=1.8$
up to the largest considered $\beta$-values, the solid curve in the figure.
We obtain a determination of the
mass-gap of the theory in units of the
three-loop lattice cut-off parameter
\begin{equation}
{m_0}=79.62 \pm 1.92~\Lambda_{latt}.
\end{equation}
The quoted error is of statistical nature.
This result agrees
with
the analytical result $m_0=80.0864 \Lambda_{latt}$ derived from the
thermodynamic Bethe Ansatz.

\section{Conclusion}

  The calculation of the mass-gap in the asymptotically free
D=2 nonlinear sigma model from the numerical evaluation of the
path integral poses a nontrivial problem. In the standard
formulation of the theory asymptotic scaling is only exhibited
for values of the nearest neighbor coupling $\beta>2.5$. Direct
correlation function measurements of the mass-gap are
ruled out. A theoretical device is needed in order to connect
the small $\beta$-region of the theory with the perturbative regime.

  In this paper we argue, that the consideration of twisted spin
configurations in union with a finite size scaling analysis
is able to bridge the gap inbetween the small and large $\beta$-regions
of the theory. It is possible to determine the coupling parameter
flow of the spin stiffness correlation length in a $\beta$-interval
which connects both regions. Focussing on the theories spin stiffness
allows us in addition to adjust certain parameters
of the finite size scaling law to their perturbative values. These are obtained
either in the numerical simulation or can be given by theoretical arguments.
The measured asymptotic value of the spin stiffness correlation length
serves as an integration constant for the intgration of $\xi_s$ towards
lower $\beta$-values, while the spin stiffness constant $\rho_s$
adjusts the amplitude of single logarithmic scaling terms of the considered scaling law.
This facilitates the precise extraction of the
$\Delta \beta (\beta)$-shift corresponding to the nearest neighbor
coupling dependence of the stiffness correlation length.
Our analysis strongly supports the existence of a scaling window
starting at $\beta$-values of about $1.8$ with an extension
into the perturbative scaling regime. Within this region the mass-gap
and stiffness correlation length are presumed to flow in coupling parameter space
under the rule of a universal
$\Delta \beta (\beta)$-shift. The integration of the shift in union
with a start value for the mass-gap confirms the known theoretical mass-gap result.
During the course of our study we have not found any indication
in favor of
the existence of a phase with vanishing mass-gap in the sigma model.
It would be interesting to apply the presented method to twisted
gauge field configurations and calculate the mass-gap of lattice QCD.

{\bf Acknowledgments:}
The author likes to thank A. Billoire for helpful comments and
M. G\"ockeler for many
discussions on the constraint effective potential.
\hfill\break

\vfill\eject

\vfill\eject

\section*{Tables}

\begin{table}[h]
\begin{center}
\begin{tabular}{||c|c||}\hline
 $\beta$ & $\Delta \beta(\beta)$ \\ \hline
        1.8  &               .10055(100) \\ \hline
        1.9  &               .10233(051) \\ \hline
        2.0  &               .10598(053) \\ \hline
        2.1  &               .10966(048) \\ \hline
        2.2  &               .11268(044) \\ \hline
        2.3  &               .11486(042) \\ \hline
        2.4  &               .11624(040) \\ \hline
        2.5  &               .11697(038) \\ \hline
        2.6  &               .11724(035) \\ \hline
        2.7  &               .11719(034) \\ \hline
        2.8  &               .11693(034) \\ \hline
        2.9  &               .11655(037) \\ \hline
        3.0  &               .11612(041) \\ \hline
        3.1  &               .11568(046) \\ \hline
\end{tabular}
\end{center}
\caption[tab1]{The $\Delta \beta (\beta)$-shift corresponding
to
changes of the coupling $\beta$ to $\hat{\beta}=\beta-\Delta\beta(\beta)$
under the decrease of the stiffness correlation length by a factor
of $2$ : $\xi_s(\hat{\beta})={1\over2}\xi_s (\beta)$.
}
\end{table}

\vfill\hfill\eject

\section*{Figure Captions}

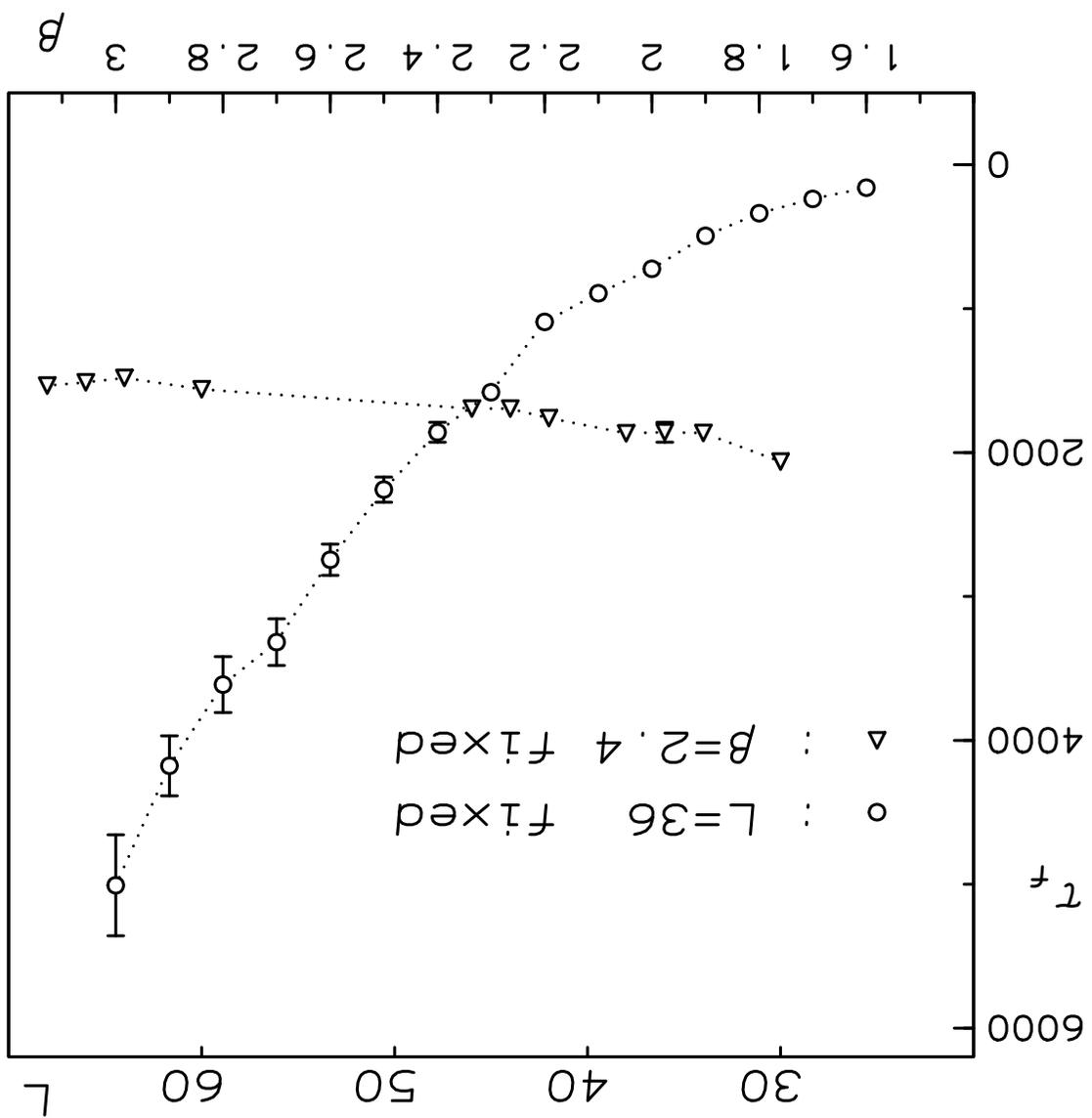
{\bf ~~~Figure 1:} Flip-autocorrelation time  $\tau_f$ of the Multicanonical Ensemble simulation
as defined in section 3 for a fixed $L=36$ lattice as a function
of $\beta$ (circles), and at fixed $\beta=2.4$ as a function
of the linear extent $L$ (triangles).

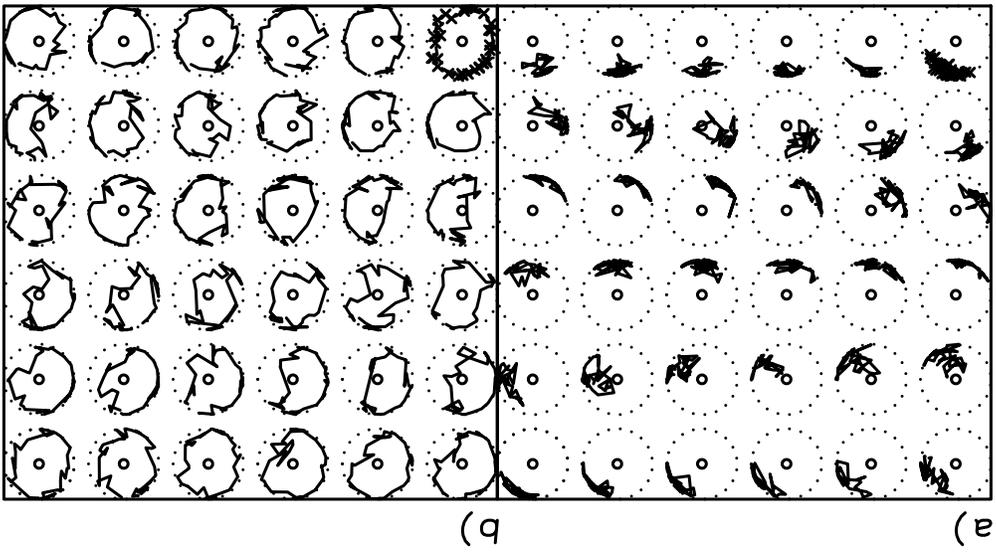
{\bf Figure 2:} Graphical representation of a field configuration
on a $36^2$ lattice at vanishing mean field $\bar\Phi$. The fields
on paths are mapped onto the complex plane and neighbors within a path
are connected by lines. In Fig. 2a) $36$ such mappings are plotted within
their corresponding unit circles, if the paths are chosen to be the main
axis of the lattice in the 1-direction. Fig. 2b) contains the
corresponding mappings, if the paths are chosen in the 2-direction.
A winding angle of $2\pi$ is observed in the latter case.
The $\beta$-value is $\beta=8$.

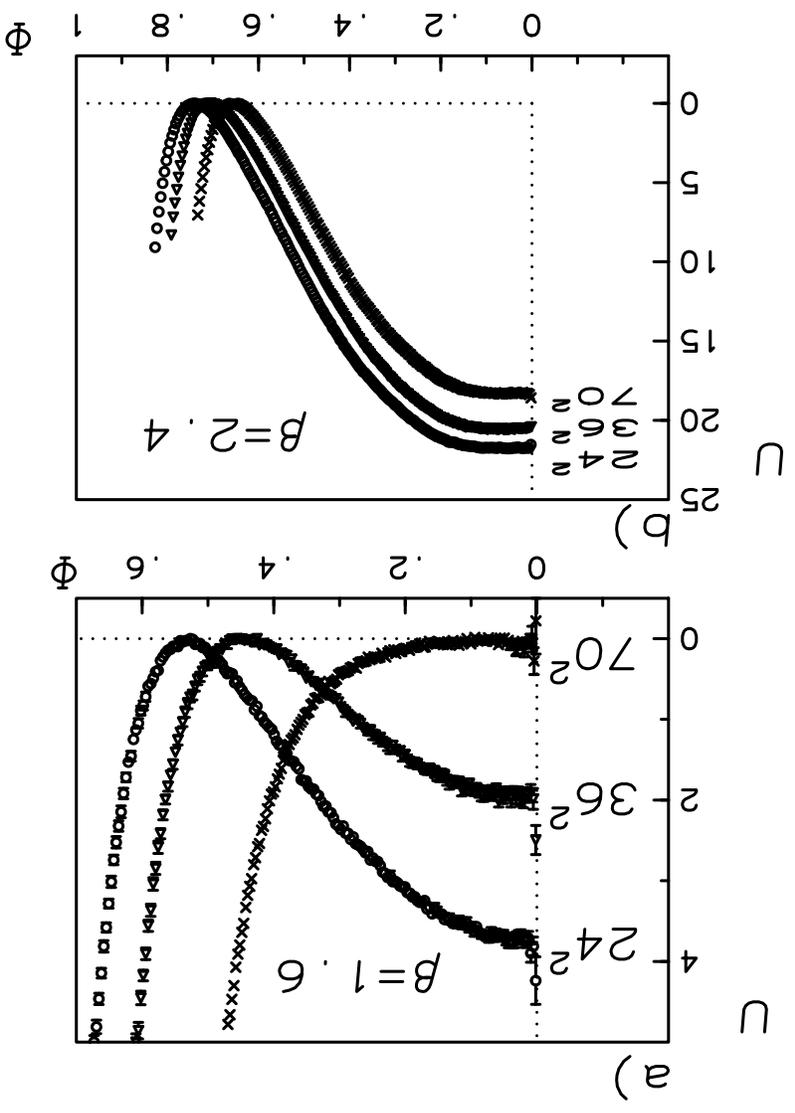
{\bf Figure 3:} The CEP on lattices of
size $L=24$ (circles), $L=36$ (triangles) and $L=70$ (crosses)
at $\beta=1.6$ (Fig. 3a)) and at $\beta=2.4$ (Fig. 3b)).

{\bf Figure 4:} The quantity $\Delta U-2\pi^2\beta$ on a
$36^2$ lattice as function
of the coupling $\beta$. The inlay displays the quantity $\Delta U$
without subtraction. The solid curve corresponds to the fit
as described in section 4 of the paper. For reasons of simplicity we omit
the drawing of error bars, except for one data point.

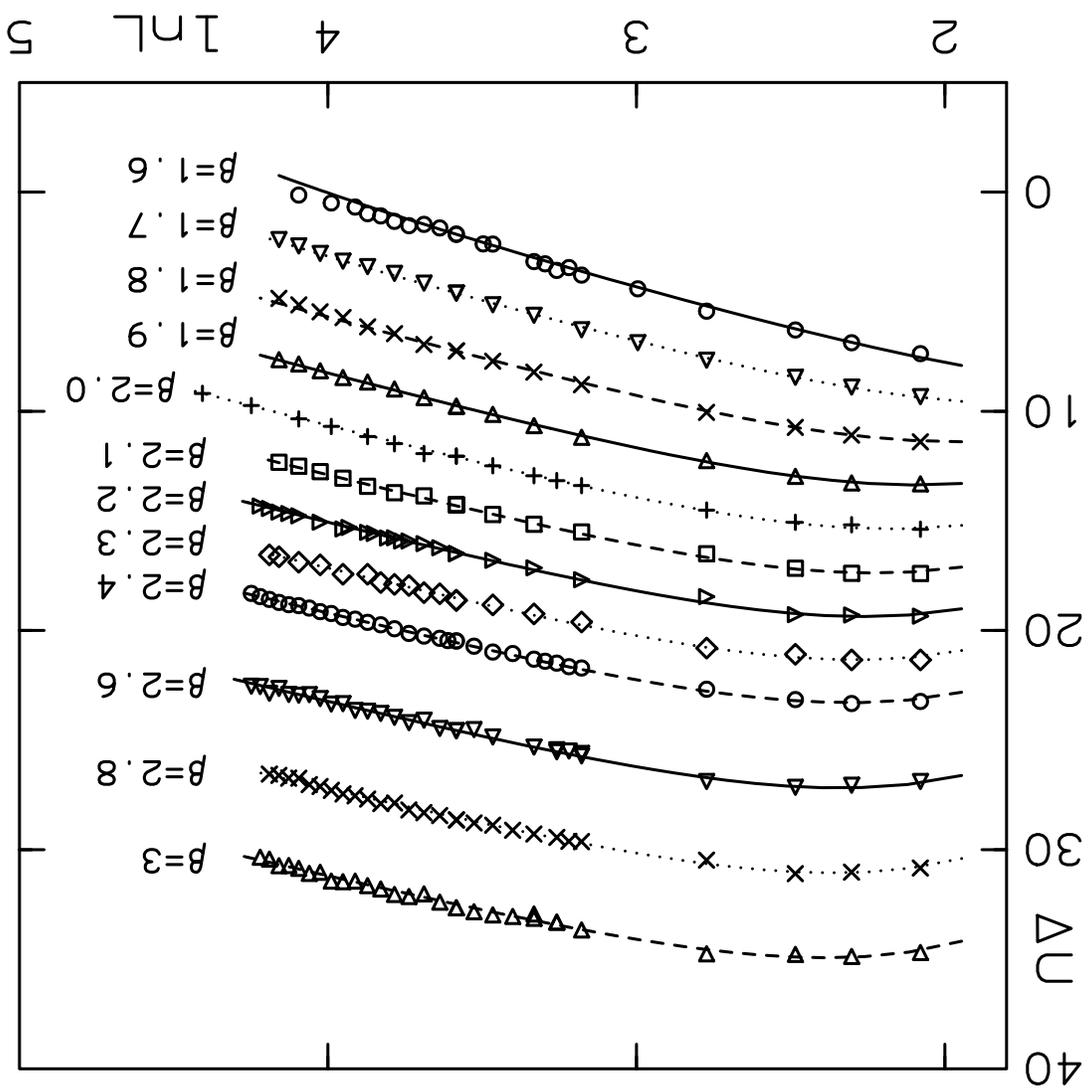
{\bf Figure 5:} The quantity $\Delta U$ as a function of $lnL$
for $12$ selected values of the coupling $\beta$.
The various curves corresponds to the fit
as described in section 4 of the paper. Note that the data
at $\beta=1.6$ have not been included into the fit. In fact small
deviations of the data from the fit become visible for the larger
lattice sizes at this value of $\beta$. For these data the
control parameter $x_s$ is not large enough.

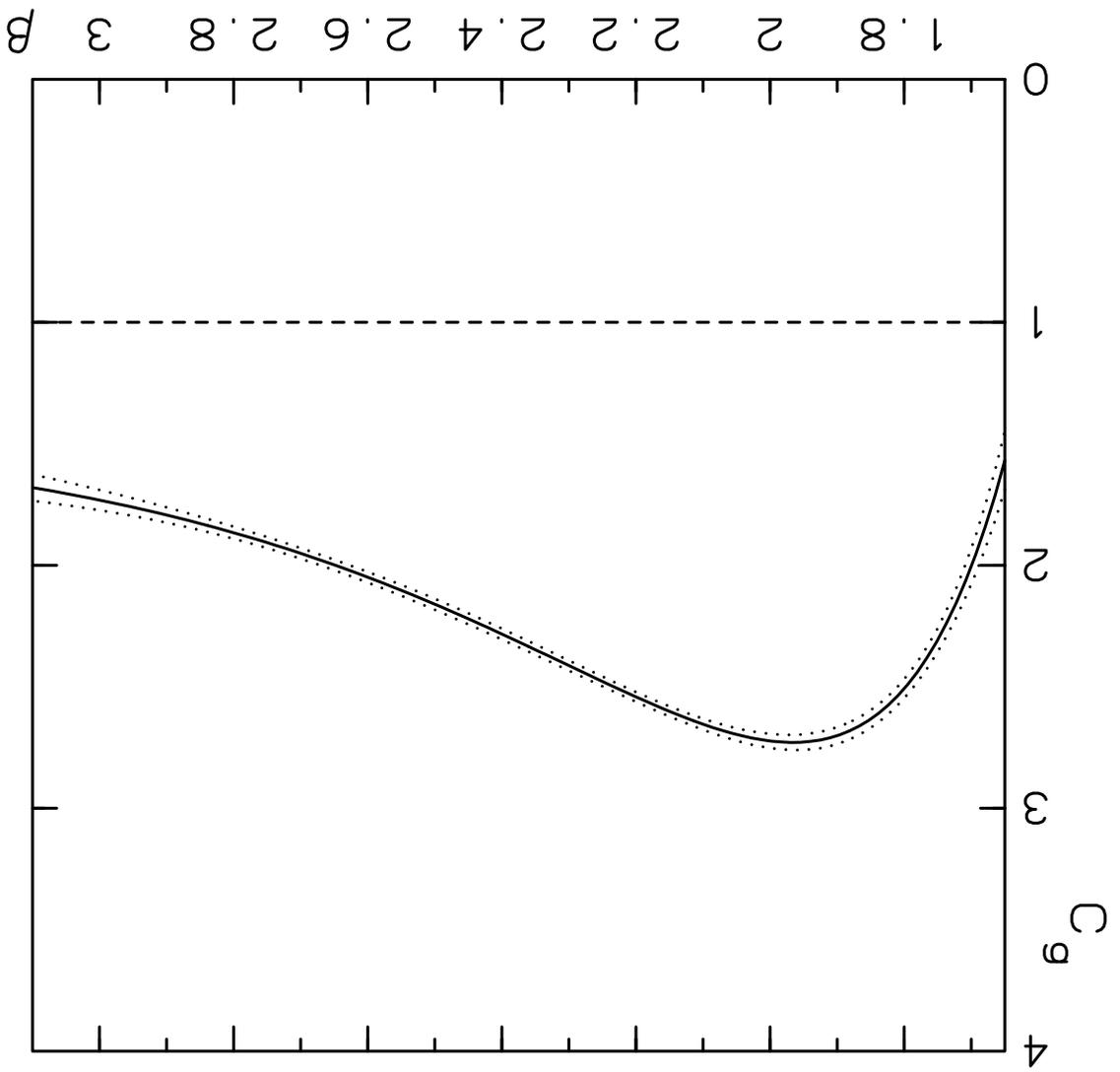
{\bf Figure 6:} The quantity $C_{g}$ as a function of
$\beta$, the solid curve. The dotted curves indicate the
error intervals as supported by the analysis. The dashed horizontal line
represents the classical limit.

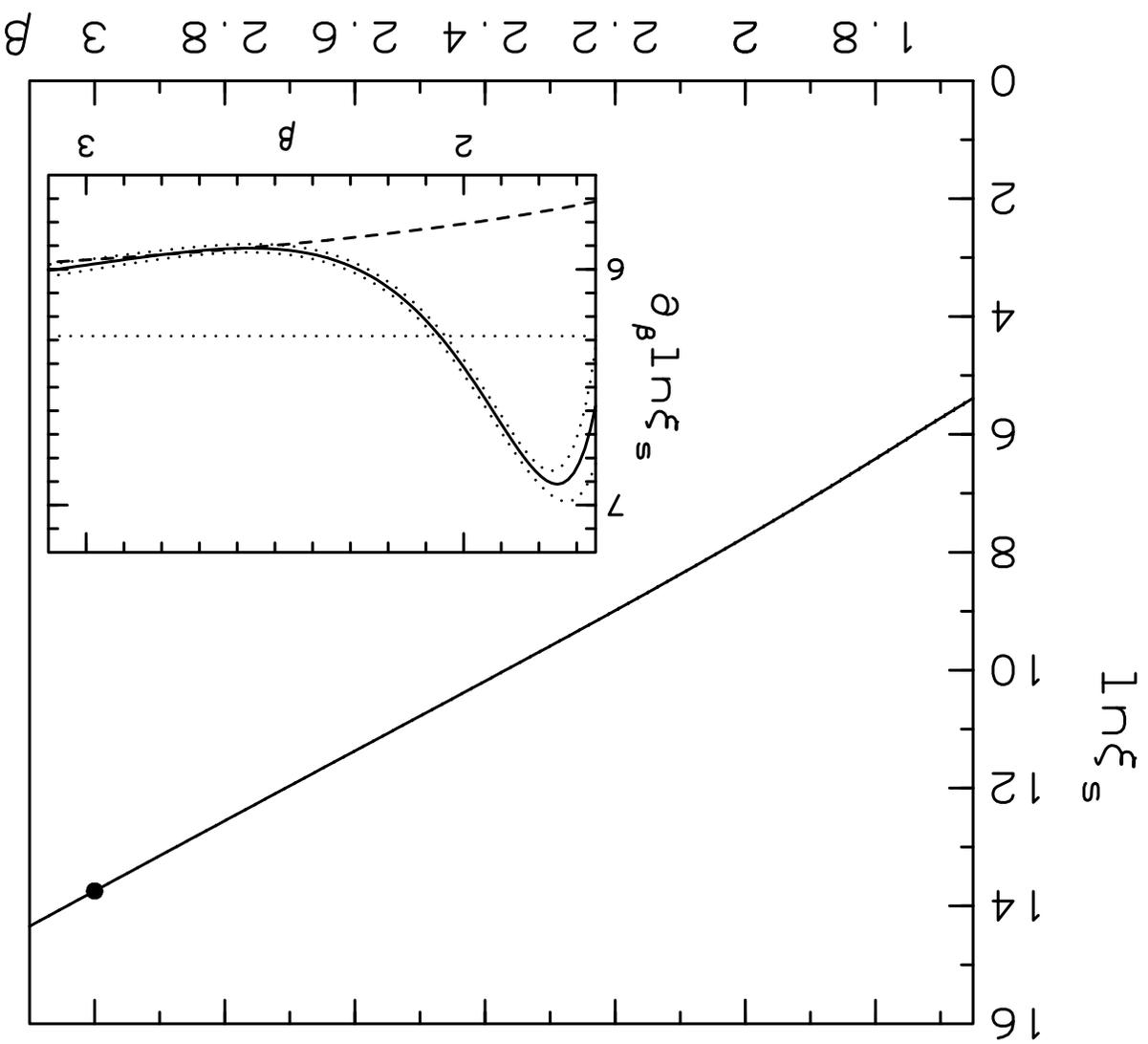
{\bf Figure 7:} Logarithmic plot of the stiffness correlation
length $\xi_s$, as obtained in section 4 of the paper. The fat circle
corresponds to the start value for $\xi_s$ in the asymptotic scaling
region. The inlay exhibits the $\beta$-derivative of $ln\xi_s$
characterized by the solid curve. Error intervals are indicated by
dotted curves. The dashed curve of the inlay represents the 3-loop
perturbative result. The inlay also contains a dotted horizontal
line at a value of $2\pi$, the constant term in the inverse
$\beta$-expansion of the considered quantity.

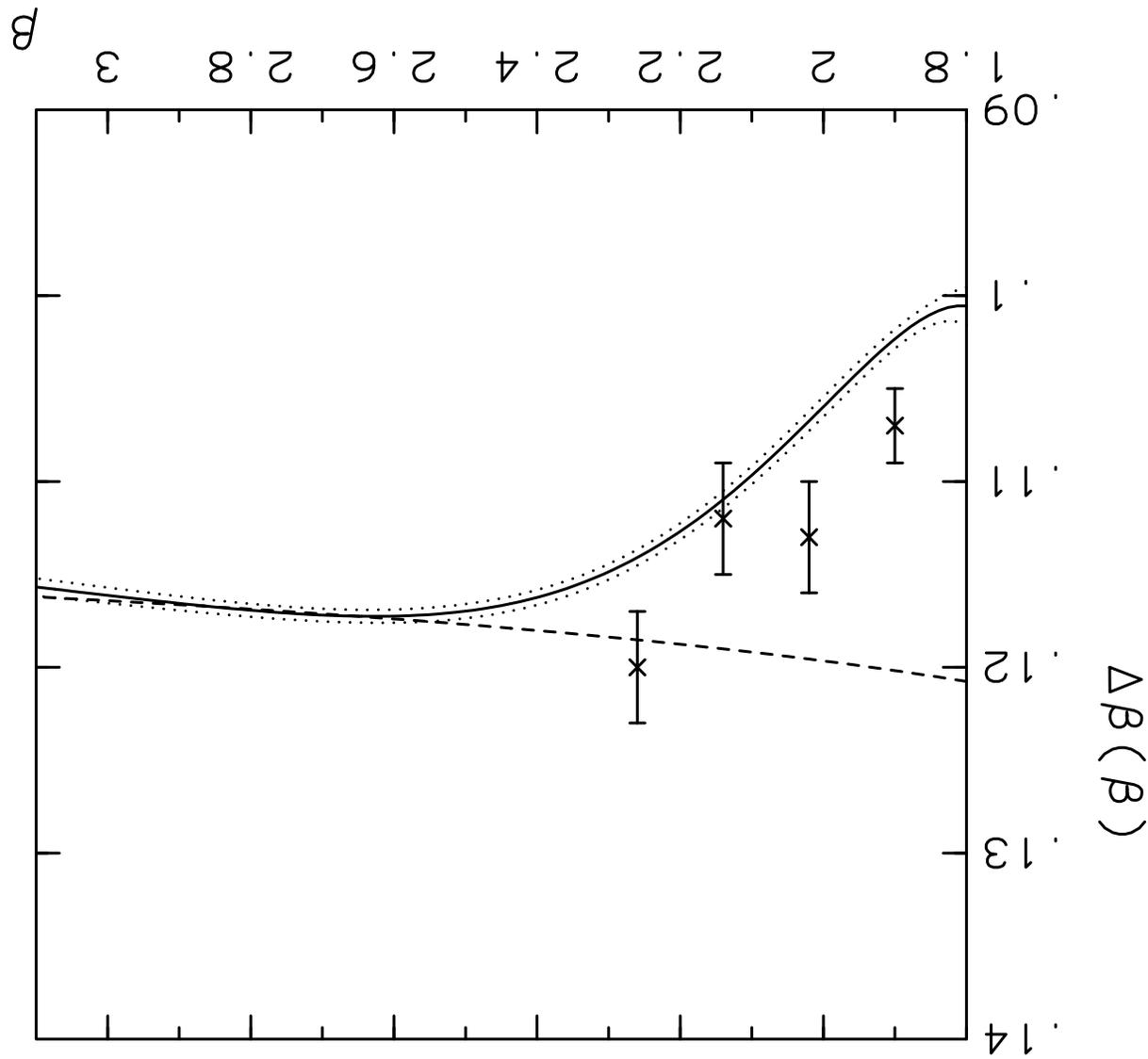
{\bf Figure 8:} The $\Delta \beta (\beta)$-shift of the spin stiffness
correlation length (solid curve) for values of $\beta>1.8$.
The dashed curve again corresponds to the 3-loop perturbative result.
Errors intervals are again indicated by dotted curves. The crosses
correspond to numerical results of Hasenfratz and Niedermayer, obtained
by a real space renormalization group matching procedure.

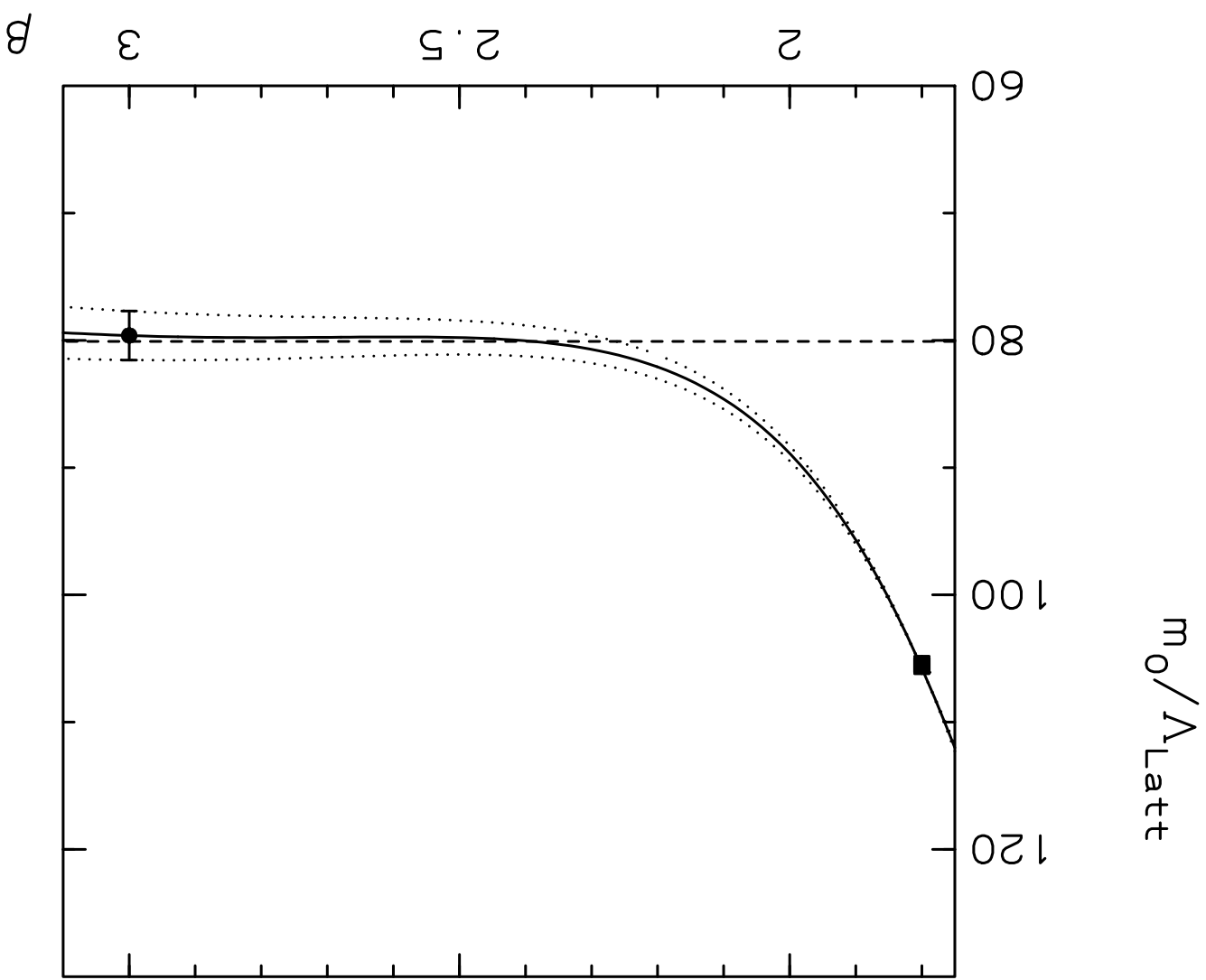
{\bf Figure 9:} The integration of the mass-gap in units of the lattice cut-off
for values of $\beta>1.8$, the solid curve in the plot. Error intervals
are treated as before. The dashed hroizontal line corresponds the exact result.
The start value for the integration is indicated by the
solid square. The asymptotic value of the mass-gap is indicated by the solid circle.

\hfill\break
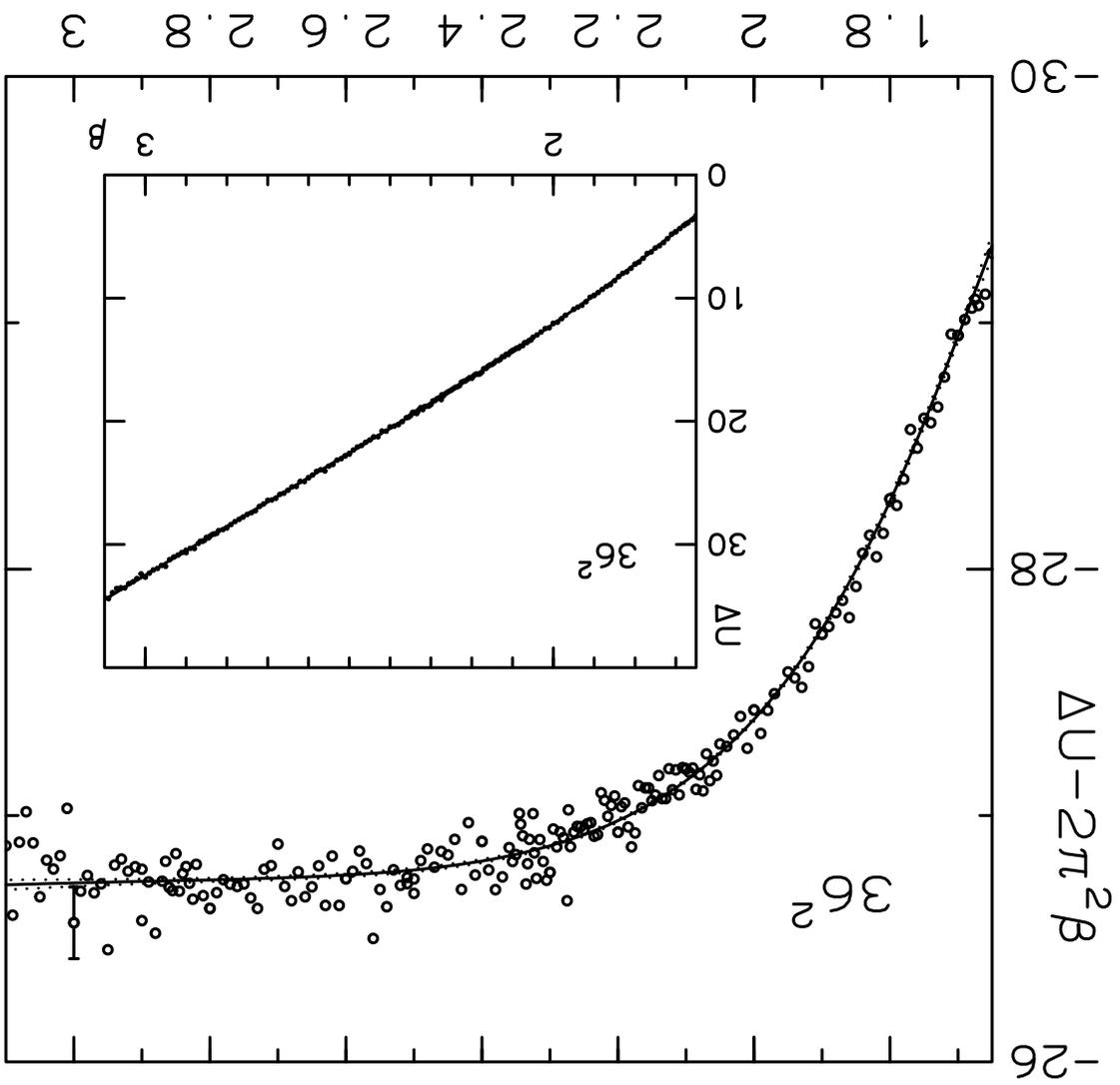
\end{document}